# Observation of Surface-Avoiding Waves: A New Class of Extended States in Periodic Media

M. Trigo,[1] T. A. Eckhause,[1] M. Reason,[2] R. S. Goldman,[2] and R. Merlin[1]

[1]*FOCUS Center and Department of Physics, The University of Michigan,*
*Ann Arbor, MI 48109-1040*

[2]*Department of Materials Science and Engineering, The University of Michigan,*
*Ann Arbor, MI 48109-2136*

Coherent time-domain optical experiments on GaAs-AlAs superlattices reveal the existence of an unusually long-lived acoustic mode at ~ 0.6 THz, which couples weakly to the environment by evading the sample boundaries. Classical as well as quantum states that steer clear of surfaces are generally shown to occur in the spectrum of periodic structures, for most boundary conditions. These surface-avoiding waves are associated with frequencies outside forbidden gaps and wavevectors in the vicinity of the center and edge of the Brillouin zone. Possible consequences for surface science and resonant cavity applications are discussed.

The behavior of waves in periodic structures has attracted the interest of scientists for many centuries [1]. Early studies go back to the late 1600's when Newton solved the problem of a one-dimensional lattice of masses connected by springs to derive an expression for the velocity of sound in air [2]. The characteristics of the solutions to wave equations are discussed in many optics, acoustics, and solid state textbooks; see, *e. g.*, Refs. [3,4,5,6]. For an infinite lattice, periodicity leads to gaps in the frequency spectrum and extended eigenmodes that satisfy the Bloch theorem, whereas isolated defects, including surfaces, may lead to localized states, such as those proposed by Tamm [7], which decay away exponentially into the bulk. In this Letter, we show that extended waves near the center and edge of the Brillouin zone have a tendency to avoid the boundaries, irrespective of the boundary conditions. Despite the long history of the problem of periodic media, this property does not seem to have been recognized before (however, see [8]). We also report time-domain optical experiments on the propagation of THz sound in a GaAs-AlAs periodic superlattice (SL). Consistent with the theoretical results, we identify a near-zone-center mode whose lifetime is abnormally long because it avoids the interface with the substrate. Although surface avoidance is a phenomenon inherent to all (classical or quantum) waves, the discussion below focuses mainly on the behavior of sound in layered media, particularly longitudinal acoustic (LA) modes, to facilitate comparison with experiments.

We studied a 75 period GaAs-AlAs SL consisting of alternating layers of thicknesses $d_{\text{GaAs}} = 59$ Å and $d_{\text{AlAs}} = 23.5$ Å, grown by molecular beam epitaxy on a (100) GaAs substrate. A 70-nm-thick layer of AlAs was deposited between the SL and the substrate to act as an etch stop. Coherent acoustic modes were generated using ~ 70 fs laser pulses provided by an optical parametric amplifier, tunable in the range 400-650 nm, at the repetition rate of 250 kHz. There are two main processes by which light couples to sound, namely, through the strain-modulation of the optical constants, as described by the photoelastic tensor [9], and through thermal stress induced by light absorption [10]. When activated by sufficiently fast pulsed sources, the latter mechanism leads to the generation of a coherent acoustic wavepacket with a size determined by the light absorption depth [10,11]. In layered media, the wavevector-selective acousto-optic mechanism accounts for spontaneous [12,13] and stimulated [14,15] Raman scattering by folded phonons (FP). Our experiments were performed at room temperature using a standard pump-probe setup in the reflection geometry, in which the only phonons that can be generated or scattered are LA modes propagating along the *z*-axis [001] [14]. The pump pulses generate coherent vibrations that induce changes in the optical constants and, thereby, scatter the weaker probe pulses that follow behind [14]. The laser beams, with an average power of 16 mW (pump) and 1 mW (probe), were linearly polarized along [110] and were focused onto nearly overlapping 10-μm-diameter spots. We measured the differential reflectivity $\Delta R \propto |\mathbf{E}_R + \Delta \mathbf{E}_R|^2 - |\mathbf{E}_R|^2 \approx 2\mathbf{E}_R \cdot \Delta \mathbf{E}_R$ as a function of the time delay between the two pulses ($\Delta \mathbf{E}_R$ is the pump-induced change in the electric field of the reflected probe beam, $\mathbf{E}_R$). Irrespective of the generation mechanism, scattering is determined by the photoelastic modulation [16]. The *bulk* component of the differential reflectivity, given by [10,14]

$$\Delta R_B \propto \int e^{i(k_I \pm k_S)z} \sigma(z) P(z) dz \qquad , \qquad (1)$$

is proportional to *L*, the total thickness of the SL. Here $\sigma = \partial u / \partial z$ is the uniaxial stress associated with the LA strain $u(z)$, $P$ is the appropriate photoelastic constant and $k_I$ ($k_S$) is the magnitude of the wavevector of the incident (scattered) light. It follows that $\Delta R_B$ is non-zero only for modes with $q = |(k_I \pm k_S) + 2\pi p/\ell|$ ($\ell = 82.5$ Å is the SL period and $p$ is an integer). These wavevectors, corresponding to $q_{BS} \approx 4\pi n(\lambda_C)/\lambda_C$ and $q_{FS} = \Omega/c \approx 0$ in the reduced-zone scheme, satisfy phase-matching conditions (quasi-momentum conservation) for back- and forward-scattering, respectively. Here $n$ is the refractive index, $\Omega$ is the LA phonon frequency and $\lambda_C$ the central wavelength of the light pulses. As described later, both back- and forward-scattered contributions are relevant to our experiments. We note that, although modes at $q_{BS}$ and $q_{FS}$ have been previously identified in pump-probe experiments on FPs [17,18,19], the generation mechanism of the zone-center mode has remained elusive. In addition to the bulk term, Eq. (1), scattering at the SL-vacuum interface leads to an additional, length-independent *boundary* component $\propto |\frac{\partial \varepsilon}{\partial u} u(t)|$ [14], where $\varepsilon$ is the dielectric function. Since the incident and scattered probe pulses are collinear and the $q_{FS}$-contribution to Eq. (1) relies on light reflected at the SL-substrate interface, it is clear that $q_{BS}$-modes dominate the bulk term. It can also be shown that only forward-scattered modes contribute to the boundary component.

The calculated spectrum of LA modes for our sample and a few representative eigenvectors are shown in Fig. 1. The displacement pattern $u(z)$ and the eigenfrequencies were obtained numerically from the wave equation

$$\frac{\partial}{\partial z}\left[C_{11}(z)\frac{\partial u}{\partial z}\right] - \rho(z)\frac{\partial^2 u}{\partial t^2} = 0 \qquad (2)$$

where ρ is the density and $C_{11}$ is the stiffness constant for LA modes propagating along the *z*-axis. Since values for GaAs and AlAs are very similar, $|\Delta C_{11}/C_{11}| \sim 0.014$ [20], we can approximate $C_{11}(z) \approx \tilde{C}_{11} \equiv \left( d_{GaAs} C_{11}^{GaAs} + d_{AlAs} C_{11}^{AlAs} \right)/\ell$. The figure draws attention to (*i*) two modes of wavevector $q_{BS}$, which are associated with the ubiquitous FP-doublets observed in backscattering Raman spectra [12,13], (*ii*) a gap mode (GM) and (*iii*) a mode close to, but not exactly at $q = 0$ whose amplitude diminishes towards both the SL-vacuum and SL-substrate interfaces. We recall that the low-frequency zone-center mode possesses the even symmetry $A_1$ and is, therefore, Raman allowed in the forward scattering geometry [21] whereas its higher-frequency counterpart does not scatter light because its symmetry, $B_2$, is odd [12,13]. Furthermore, since the dispersion at $q = q_{BS}$ is nearly linear, the FP modes behave as weakly-modulated plane waves. The most important feature of Fig. 1 is the boundary-avoiding character of the near-zone-center eigenfunction. The calculations show that all SL eigenmodes associated with wavevectors in close proximity to the center and edge of the Brillouin zone shy away from the boundaries and, in view of that, we refer to them as surface-avoiding modes (SAMs). Below, we show that surface avoidance is the general solution to the problem of satisfying certain boundary conditions with almost-standing waves.

For a periodic structure, Bloch theorem dictates that the extended solutions are of the form $u(z,t) = U_{q,s}(z) e^{i(qz - \Omega t)}$ where $U_{q,s}$ is periodic (*s* is the branch or band index). From (2) we obtain

$$\frac{\partial^2 U_{q,s}}{\partial z^2} + 2iq \frac{\partial U_{q,s}}{\partial z} + \left[ \frac{\rho(z) \Omega_s^2}{\tilde{C}_{11}} - q^2 \right] U_{q,s} = 0 \qquad . \qquad (3)$$

For $|q| \ll \pi/\ell$, we approximate $U_{q,s} \approx U_s^{(0)} + q\ell U_s^{(1)}$ and $\Omega_s(q) \approx \Omega_s(0)$. Assuming that the functions are normalized as $\int \rho(z) U_{0,s}^* U_{0,r} dz = \delta_{sr}$, we use perturbation theory to get

$U_s^{(0)} = U_{0,s}$ and $U_s^{(1)} = \dfrac{2i\tilde{C}_{11}}{\ell} \sum_r \dfrac{\int U_{0,r}^* \left(\partial U_{0,s}/\partial z\right) dz}{(\omega_{0,s}^2 - \omega_{0,r}^2)} U_{0,r}$. Consider a semi-infinite SL occupying the half-space $z > 0$ and assume that $u + \beta \partial u/\partial z = 0$ at $z = 0$. This boundary condition covers the free ($\beta \to \infty$), clamped ($\beta = 0$) and all intermediate situations for which the waves are non-propagating in the half-space $z < 0$, so that Im($\beta$)=0. It also includes the problem of reflection (or scattering), corresponding to Re($\beta$) = 0, when sound is incident from the SL side, but not when the waves originate from the $z < 0$ region of the interface. Ignoring the band index and the factor exp(-$i\Omega t$), the solutions for $z > 0$ are of the form $u(z) = (A_+ U_{+q} e^{iqz} + A_- U_{-q} e^{-iqz})$. To lowest order, the displacement is

$$u(z) \approx -iU^{(0)}(z)\left[U^{(0)}(0) + \beta \dot{U}^{(0)}(0)\right] \sin qz +$$
$$q\ell \left\{ \begin{array}{l} U^{(0)}(z)\left[i\dfrac{\beta}{\ell} U^{(0)}(0) + U^{(1)}(0) + \beta \dot{U}^{(1)}(0)\right] \\ -U^{(1)}(z)\left[U^{(0)}(0) + \beta \dot{U}^{(0)}(0)\right] \end{array} \right\} \cos qz \quad . \quad (4)$$

As expected, this expression involves the product of rapidly- ($U_s^{(0)}$, $U_s^{(1)}$) and slowly- ($\sin qz$, $\cos qz$) varying functions, and shows beats due to the interference between terms such as $(2\pi/\ell) - q$ and $(2\pi/\ell) + q$. More importantly, it describes a *surface-avoiding* mode. The boundary-repelling character of the solution becomes evident if we compare the amplitudes at $z = 0$ and $z \approx \pi/2q$ which, except for the trivial case $U^{(0)}(0) + \beta \dot{U}^{(0)}(0) = 0$, are in a ratio $\propto q\ell \ll 1$. It is apparent that such a behavior is due primarily to the fact that $U_{+q}$ and $U_{-q}$ cease to be linearly independent at $|q| = 0$ [22].

Some thought shows that the above arguments apply to all linear wave equations. In particular, we note that the **k.p** method of electron band structure gives an expression for quantum states that is identical to Eq. (4). Hence, for the boundary condition we have chosen, the near zone-center and zone-edge solutions to Schrödinger's and Maxwell's equations are necessarily SAMs (however, note that surface-avoidance is not required if the source of waves is *outside* the periodic medium). It should also be noted that surface-avoidance does not follow from the fact that, at integer multiples of $\pi/\ell$, the physically-sound solutions in the infinite limit are standing waves [22]. While all the eigenstates for a finite and acoustically-isolated sample are standing waves, only those with frequencies close to the zone-center and zone-edge frequencies are SAMs.

While these results are interesting in their own right, we believe that they also have practical implications, especially for work on low-dimensional systems using surface-sensitive techniques, since restricted dimensionality favors singularities in the density-of-states where SAMs occur. Furthermore, as illustrated in Fig. 1, surface-avoidance leads to behavior of the slowly-varying envelope that is reminiscent of a resonant cavity. It is interesting to note that in the SAM resonator, the SL, a Bragg-reflector typically used as a mirror in optical cavities [23], is instead the cavity itself. Since the smallest wavevector allowed for a finite SL is $\sim \pi/L$, the finesse of such a cavity is thus proportional to the number of cells. This differs significantly from a Fabry-Perot resonator for which the cavity size determines mainly the resonant wavelength (see, *e. g.*, [24] for acoustic cavities).

The results of our experiments, summarized in Figs. 2 and 3, provide strong support for the existence of surface-avoiding waves. Pump-probe data were acquired using pulse energies tuned to resonate with the SL bandgap at ~ 2.27 eV. A typical time-domain trace is reproduced in Fig. 2 whereas Fig. 3 shows the associated Fourier transforms for various time windows. Scans similar to that in Fig. 2 were obtained at delays as large as 800 ps. As shown in Fig. 3, the oscillations involve four distinct modes [25]. The dominant *B*-type oscillations are ascribed to stimulated Brillouin backscattering. The measured period, ~ 14 ps, is in excellent agreement with that of the mode of frequency $(\tilde{C}_{11}/\tilde{\rho})^{1/2} q_{BS}$ where $\tilde{\rho}$ is the average density of the SL. The higher-frequency oscillations at ~ 0.6 THz, underlined in the inset of Fig. 2, involve a triplet consisting of the FP-doublet and a third peak, at a frequency slightly shifted from the mean frequency of the doublet, which we assign to forward scattering by a near-zone-center mode. The comparison with the calculated LA dispersion, shown in Fig. 3, corroborates these assignments. We emphasize that coherent scattering due to all these modes is phase-matched and that all these features have been previously observed in spontaneous Raman and Brillouin scattering experiments [12,13]. What is crucial in Fig. 3 is the behavior of the $q \approx 0$ mode, whose presence in the time-window spectra significantly exceeds that of the other features. While the FP-doublet is very weak and the amplitude of the *B*-mode is reduced in half already at times comparable to the transit time of a longitudinal sound pulse across the SL (~125 ps), the $q \approx 0$ oscillations persist at delays as large as ~ 0.8 ns. Based on these results, the theoretical arguments discussed previously, and the fact that the coherent phonons are generated in the SL and decay into the substrate, we attribute the unusual long lifetime of the near-zone-center mode to surface-avoidance. An upper bound for the SAM decay rate, γ, can be estimated from the time of flight needed to reach the SL-substrate interface. This gives $\gamma \sim v_G(1-\mathfrak{R})/L$, where $\mathfrak{R}$ and $v_G$ are the sound reflection coefficient and group velocity. Using that $q_{\text{SAM}} \sim \pi/L$, we obtain $\gamma^{-1} \sim 5$ ns which is consistent with our observations.

In summary, we have presented theoretical arguments as well as experiments on acoustic waves in a SL, which demonstrate the existence of a new class of states in periodic media, namely, waves that stay away from boundaries. Our results apply both to quantum and classical modes and hold promise for a wide range of experimental studies in surface physics and resonant-cavity applications.

This work was supported by the AFOSR under contract F49620-00-1-0328 through the MURI program and by the NSF Focus Physics Frontier Center.

[25] Results obtained with a Ti:Sapphire resonator operating at 800 nm (~ 0.7 eV below the SL gap) show instead a single peak corresponding to the $q \approx 0$ mode. In this case, the boundary contribution dominates because the laser is not absorbed in the SL.

# Figure Captions

FIGURE 1 (color online). Calculated dispersion of LA modes propagating along [001] for the GaAs-AlAs SL investigated in this work. Panels on the right show a schematic diagram and four representative eigenmodes of the 619-nm-thick SL slab (striped area) attached to a 5-µm-thick substrate. Vertical green bars denote the SL-substrate boundary. SAM is the surface-avoiding mode at $q \approx 0$. The wavevector associated with modes labeled FP, $q_{BS}$ (see text), is sufficiently far removed from the zone center so that the FP-waves reflect only weakly at the boundary. The mode with frequency in the minigap, GM, decays exponentially into the SL.

FIGURE 2 (color online). Differential reflectivity data at 300 K and 10 ps $< t <$ 250 ps showing coherent acoustic oscillations. The central wavelength of the laser pulses is $\lambda_C = 546$ nm. A slowly varying background has been subtracted. The yellow rectangle indicates the range of the scan shown in the inset.

FIGURE 3 (color online). Fourier transform of the time-domain data for the intervals: 10–125 ps (a), 250-350 ps (b) and 350-500 ps (c). SAM is the surface-avoiding wave at $q \approx 0$. The bottom panel shows the inverted dispersion relation (parameters are the same as in Fig. 1). The horizontal line represents the wavevector $q_{BS}$ determined from phase-matched backscattering conditions.

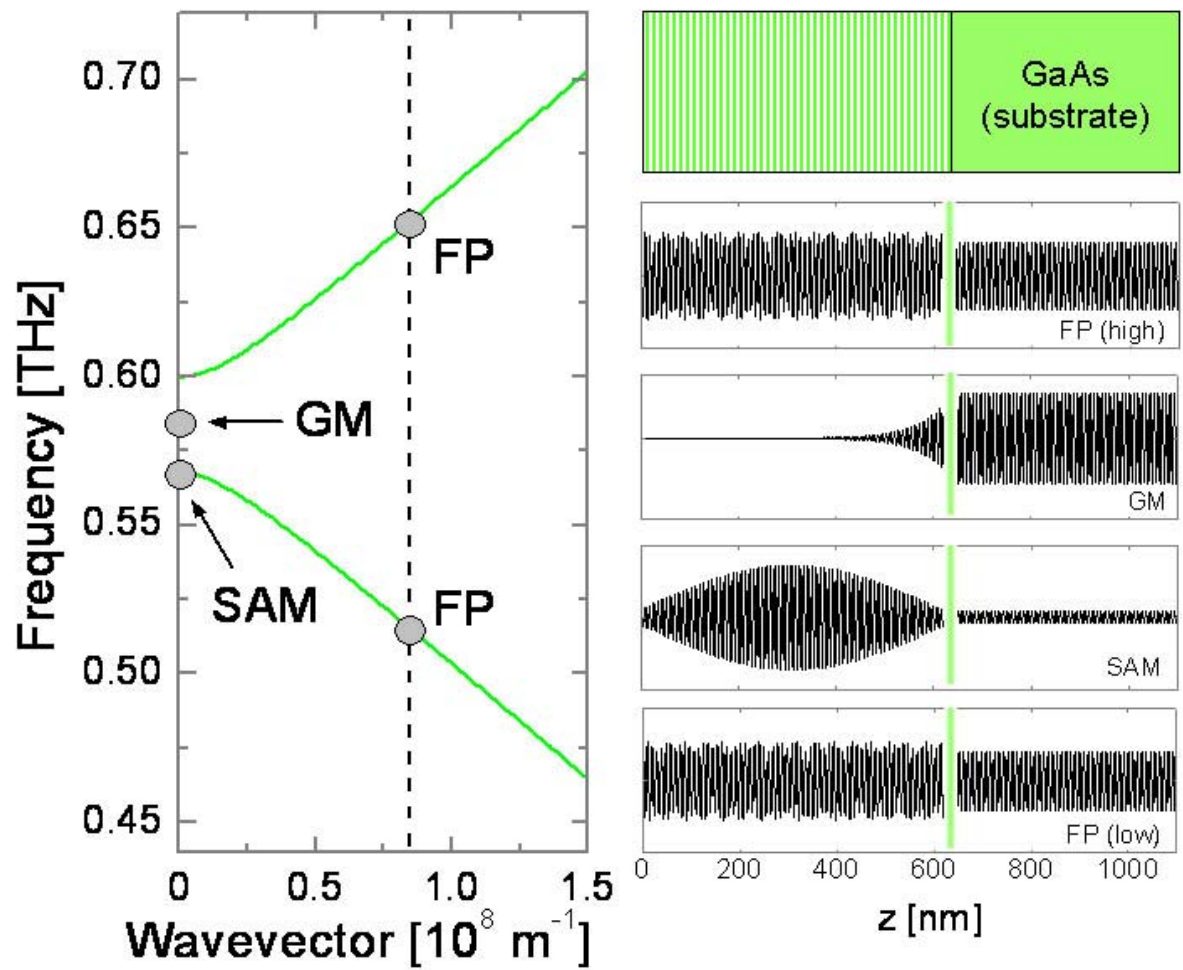

FIGURE 1

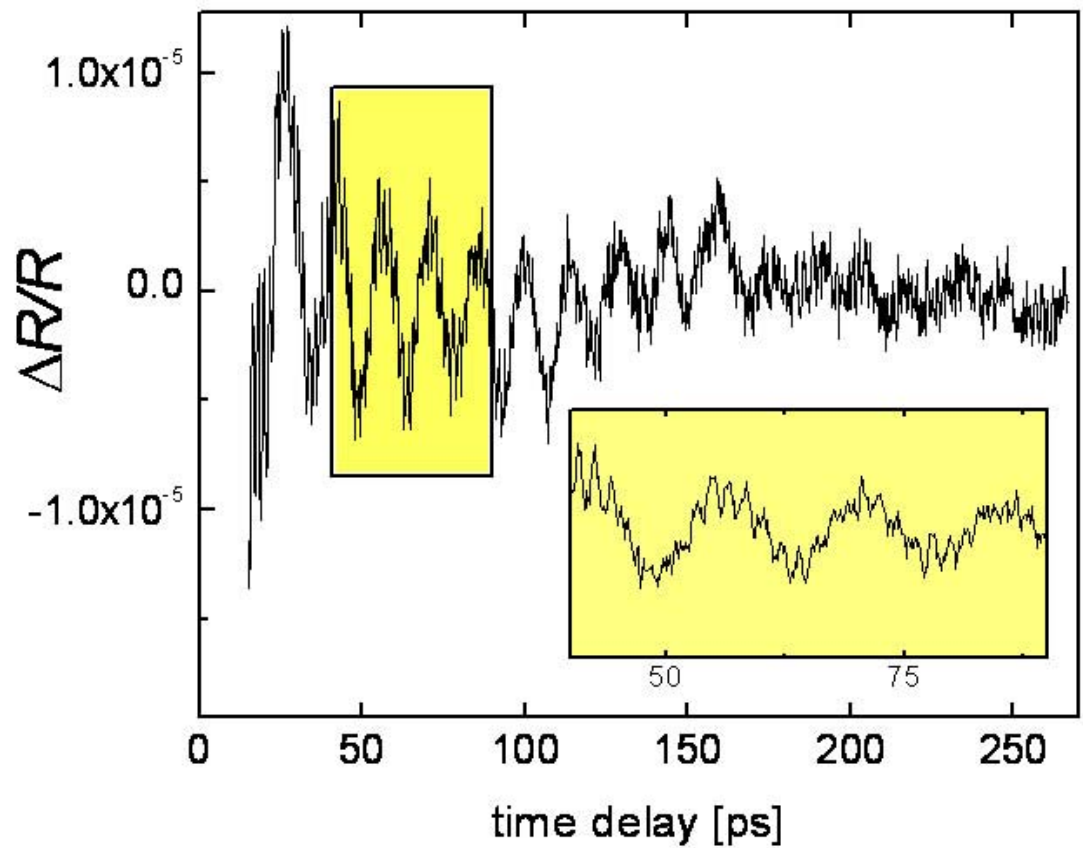

FIGURE 2

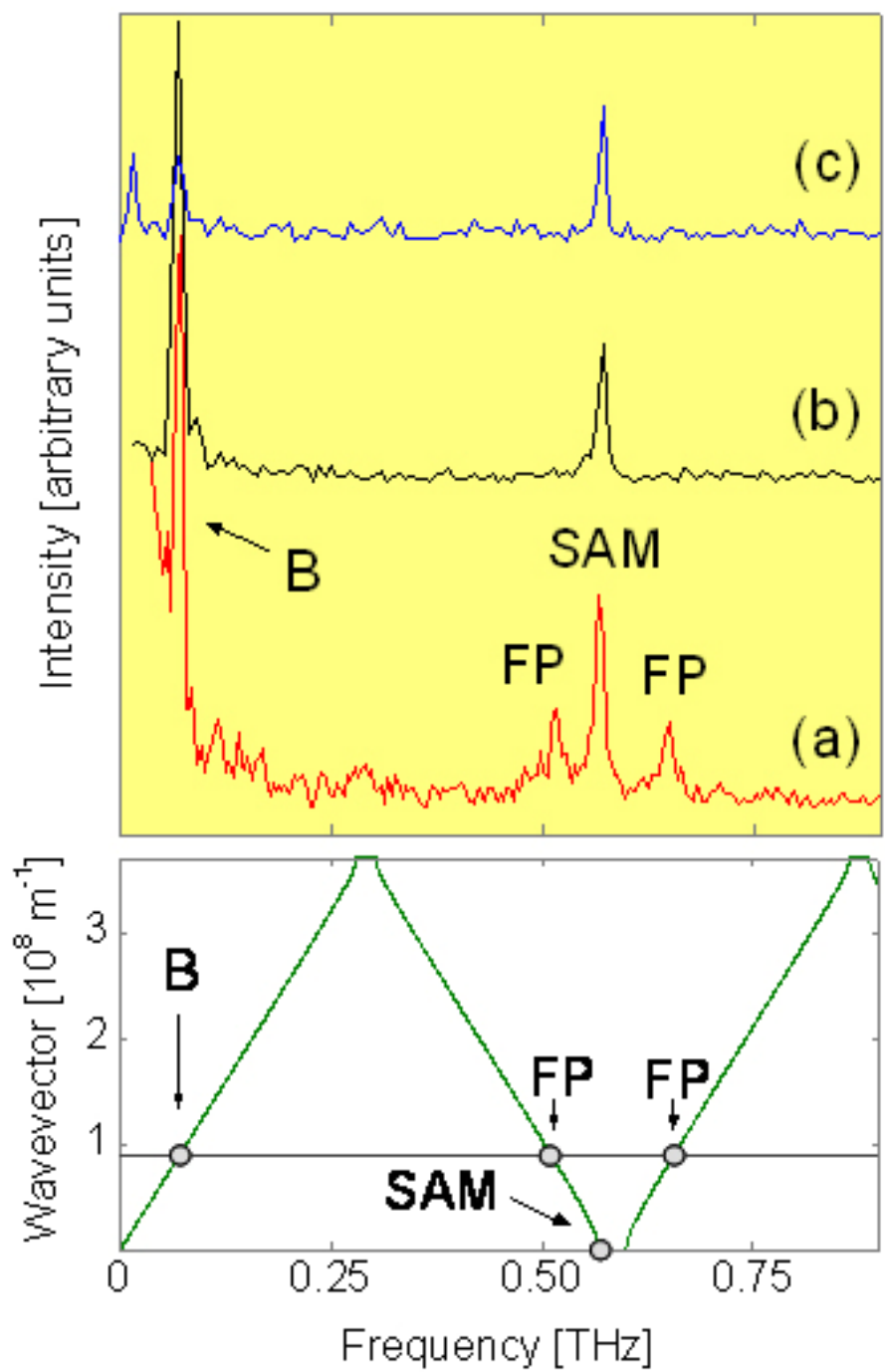

FIGURE 3